\documentclass[pdftex,12pt]{article}
\usepackage{times}
\usepackage{geometry}
\usepackage[utf8]{inputenc}
\usepackage{enumitem,amssymb}
\usepackage{ragged2e}
\usepackage[numbers]{natbib}
\usepackage{graphicx}
\usepackage{subcaption}
\newlist{thematic}{itemize}{8}
\setlist[thematic]{label=$\square$}
\usepackage{pifont}
\usepackage{wrapfig}
\newcommand{\cmark}{\ding{51}}%
\newcommand{\done}{\rlap{$\square$}{\raisebox{2pt}{\large\hspace{1pt}\cmark}}%
\hspace{-2.5pt}}

\newcommand{\colibri}{\textit{Colibr\`i}}

\setlist[itemize]{itemsep=2pt,wide=10pt,leftmargin=\dimexpr\labelwidth + 3\labelsep\relax,topsep=5pt}

\begin{document}
\thispagestyle{empty}
{\raggedright
\huge
Astro2020 Science White Paper \linebreak

Exploring the physics of neutron stars with high-resolution, high-throughput X-ray spectroscopy \linebreak
\normalsize

\noindent \textbf{Thematic Areas:} \hspace*{60pt} $\square$ Planetary Systems \hspace*{10pt} $\square$ Star and Planet Formation \hspace*{20pt}\linebreak
$\done$ Formation and Evolution of Compact Objects \hspace*{31pt} $\done$ Cosmology and Fundamental Physics \linebreak
  $\square$  Stars and Stellar Evolution \hspace*{1pt} $\square$ Resolved Stellar Populations and their Environments \hspace*{40pt} \linebreak
  $\square$    Galaxy Evolution   \hspace*{45pt} $\square$             Multi-Messenger Astronomy and Astrophysics \hspace*{65pt} \linebreak
  
\textbf{Principal Author:}

Name: Jeremy Heyl	
 \linebreak						
Institution: University of British Columbia
 \linebreak
Email: heyl@phas.ubc.ca
 \linebreak
Phone:  +16048220995
 \linebreak
 
\textbf{Co-authors:} 
Ilaria Caiazzo (UBC),
Samar Safi-Harb (Manitoba),
Craig Heinke (Alberta),
Sharon Morsink (Alberta),
Edward Cackett (Wayne State),
Alessandra De Rosa (INAF/IAPS Roma),
Marco Feroci (INAF/IAPS Roma),
Daniel S. Swetz (NIST),
Andrea Damascelli (UBC-QMI),
Pinder Dosanjh (UBC-QMI),
Sarah Gallagher (Western/CSA),
Luigi Gallo (St Mary's),
Daryl Haggard (McGill),
Kelsey Hoffman (Bishop's),
Adam R. Ingram (Oxford),
Demet K{\i}rm{\i}z{\i}bayrak (UBC),
Herman Marshall (MIT),
Wolfgang Rau (Queens/TRIUMF),
Paul Ripoche (UBC),
Gregory R. Sivakoff (Alberta),
Ingrid Stairs (UBC),
Luigi Stella (INAF - Osservatorio Astronomico di Roma),
Joel N. Ullom (NIST).
}
\\
\\
\textbf{Abstract:} The advent of moderately high-resolution X-ray spectroscopy with \textit{Chandra} and \textit{XMM} promised to usher in a new age in the study of neutron stars: we thought we would study neutron stars like stars, with resolved absorption spectra revealing their surface chemical composition and physical conditions (e.g. surface gravity, pressure, temperature). Nature, however, did not cooperate in this endeavor, as observations of neutron stars have not revealed verified atomic absorption lines yet. In the near future, advancements in transition-edge sensors (TES) technology will allow for electron-volt-resolution spectroscopy combined with nanoseconds-precision timing.  Combining these detectors with collector optics will also us to study neutron stars in much greater detail by achieving high-energy resolution with much larger collecting areas to uncover even weak spectral features over a wide range of the photon energies.  Perhaps  we will finally be able to study neutron stars like stars.

\pagebreak
\pagenumbering{arabic}
\section{Introduction}

In the fifty years since their discovery, neutron stars have never stopped puzzling and amazing astronomers. First discovered as radio pulsars, neutron stars have revealed themselves in different fashions, over all the electromagnetic spectrum and via gravitational waves. From the almost 3,000 radio pulsars detected up to now, to the radio-quiet, thermally emitting isolated neutron stars (XDINs), from the young and active magnetars with extreme magnetic fields to the old and rapidly rotating millisecond pulsars, from accreting to merging binaries; neutron-star phenomenology is rich and we have learned a lot from it. Yet, many puzzles remain, including the key question
\textbf{
\begin{itemize}[label=$\diamond$]
\item What are neutron stars made of?
\end{itemize}
}
This question has profound implications for the physics of dense matter. The density reached in a neutron star's core, several times higher than nuclear density, is not reached anywhere else in the universe at cold temperatures, let alone in our terrestrial physics labs, and therefore neutron stars represent the only laboratory available to look for the equation of state for cold, dense matter. The holy grail of neutron star observations, the mass-radius relation, if measured for several neutron stars, could put stringent constraints on the equation of state \cite{2013ApJ...765L...5S}. Mass measurements of massive neutron stars exclude a number of equations of state that predict a relatively soft dependence of pressure on density. Although a number of masses of neutron stars have been measured with high precision, especially for compact binaries, radius measurements are much harder to achieve with the precision of less than a kilometer required to put stringent constraints on the equation of state.

The advent of moderately high-resolution X-ray spectroscopy with \textit{Chandra} and \textit{XMM} promised to usher in a new age in the study of neutron stars: we thought we would study neutron stars like stars, with resolved absorption spectra revealing their surface chemical composition and physical conditions (e.g. surface gravity, pressure, temperature). Nature, however, did not cooperate in this endeavor, as high-spectral-resolution observations of neutron stars have not revealed verified atomic absorption lines yet. Still, hints of the presence of absorption lines have been detected in accreting and isolated neutron stars, and the advent of high energy-resolution spectroscopy in the X-rays could still bring the detection of narrow and weak absorption features. 

Recent advancements in transition-edge sensors (TES) present a unique opportunity to open a new window on neutron stars: high energy-resolution spectroscopy combined with high-precision timing. TES-based detectors can already achieve an energy resolution of less than an electronvolt at about 1.5 keV, and of about 2-3 eV at 5-10 keV \cite{doi:10.1063/1.4984065}. Furthermore, the arrival times of the photons can be measured to a precision of 300~ns or better with a short deadtime of less than a millisecond on a given TES array element \cite{doi:10.1063/1.4962636}. In the context of ongoing missions, with the current TES technology it is possible to achieve the timing resolution of the best timing telescope in space right now (\textit{NICER}, with a resolution of 100 ns) while reaching an energy resolution more than 40 times better than \textit{XMM-Newton} (130 eV at 6 keV), and there is still space for improvement. The Canadian Space Agency recently funded an 18-month concept study for a TES-based telescope, \colibri, that started in September, 2018. \colibri\ will be dedicated to the study of compact objects in the X-ray, and will pair TES-based detectors with collector optics to achieve a high throughput ($\sim$100kHz count-rates).

\section{High energy-resolution, high-throughput spectroscopy of the neutron star surface}


Due to the high surface gravity of neutron stars, the elements in neutron star atmospheres stratify within 30 seconds, leading to a photosphere made from only the lightest element present, typically hydrogen. In most neutron stars, some amount of accretion (from a companion star, the interstellar medium, or fallback from the supernova) will have occurred, so featureless (in the X-ray) hydrogen atmospheres are generally expected \cite[e.g.][]{2007MNRAS.375..821H}. 
    
However, a variety of elements may be present in the photospheres of neutron stars, if the neutron stars are actively accreting. 
    On rapidly spinning neutron stars, spectral lines will be spread out by the Doppler shift; therefore, detecting and measuring the energy and width of the spectral lines from a rotating neutron star would directly provide an estimate of the neutron star radius, if the spin period is known \cite{2003ApJ...582L..31O}. In general, rapid rotation may make the lines too broad to be detected. However, the narrow {\it cores} of these broad lines may be deep enough to be clearly detected with appropriate throughput and spectral resolution  \cite{2013ApJ...766...87B}.  Also, some X-ray binaries have relatively low spin (e.g. Terzan 5 X-2, \cite{2010ATel.2929....1S}), and/or very low inclination, either of which would narrow the lines sufficiently for possible detection
    \cite{2018MNRAS.475.2194Y}.
    
During active accretion episodes, the surface of the neutron star is generally not visible, as photons from the stellar surface are Comptonized by the accreting material. 
Type-I X-ray bursts and carbon superbursts represent an exception, as thermonuclear reactions on the surface of the star dramatically increase the emission from the surface itself, so it dominates the emission for a few seconds to hours. Cottam et al. \cite{2002Natur.420...51C} identified absorption lines in the sum of \textit{XMM-Newton} spectra over many Type-I X-ray bursts from EXO~0748-676, which they argued were redshifted Fe lines from the stellar surface. This particular source is now thought to be rotating rapidly \cite{2010ApJ...711L.148G}, which makes it challenging to explain the relatively narrow spectral features that they found  \cite{2010ApJ...723.1053L,2013ApJ...766...87B}. 
    
    However, several X-ray bursts since then have shown evidence for broader features, likely due to heavy nuclear burning products being mixed up to the photosphere in particularly energetic bursts \citep{2006ApJ...639.1018W}. These include observations of an likely edge around 7.5 keV in HETE J1900.1-2455 by RXTE  \citep{2017MNRAS.464L...6K}, and around 8 keV in GRS 1747-312 by RXTE \citep{2018ApJ...866...53L}. However, RXTE's spectral resolution was insufficient to clearly identify the spectral feature high-throughput TES arrays will have the spectral resolution to clearly resolve edge features such as these, and the effective area to spot them in short time periods ($\sim1$ s), allowing robust determination of the surface redshift of these bursting neutron stars. 
    
    These particular observations are hard for planned instruments such as the \textit{X-IFU} on \textit{ATHENA} \cite{2016SPIE.9905E..2FB}, because of photon pile-up, but straightforward for instruments with collector optics. The count rate during X-ray bursts will peak at about 1-10~kHz (scaling from \textit{RXTE} results \cite{2008ApJS..179..360G}) for \textit{ATHENA} and \colibri, which both plan to use TES X-ray detectors for spectroscopy. In the case of \textit{ATHENA}, if two photons arrive within 2.6~ms from each other on the same pixel, the energy of the second photon cannot be measured. The expected count rates for X-ray bursts dramatically exceed this limit, and therefore, without blocking filters to reduce the effective area to the level of \textit{XMM-Newton} or deliberate defocusing, \textit{ATHENA} cannot perform spectroscopy during X-ray bursts. On the other hand, because a mission with collecting optics such as \colibri\ builds effective area by having many collectors operating in parallel, with each collector focusing X-rays on several elements of a TES array, its nominal configuration can achieve high-resolution spectroscopy to count rates well beyond 100~kHz.
    
   Low-mass X-ray binaries during quiescent periods between outbursts also exhibit surface features. If no accretion is occurring, then the photosphere will contain only the lightest element (typically H or He), and no lines will be present. However, if accretion onto the neutron star exceeds $\dot{M}\sim10^{-13} \mathrm{M}_\odot$/year (corresponding to $L_X\sim10^{33}$ erg/s), it is likely that metals will substantially populate the photosphere. Evidence for metal features in the photosphere of the quiescent LMXB Aquila X-1 was produced by \citet{2002ApJ...577..346R} from one (of several) Chandra observations. However, Chandra's spectral resolution and low-energy calibration left the nature of this feature in some doubt. High-energy resolution TES detectors  would permit clear identification of these spectral features, likely within the context of a program to monitor the cooling of neutron star crusts after an outburst \citep{2017JApA...38...49W}, when occasional burps of accretion are often observed.

   \section{High-energy-resolution, high-time-resolution spectroscopy of the neutron star surface}

\begin{figure}[tb]
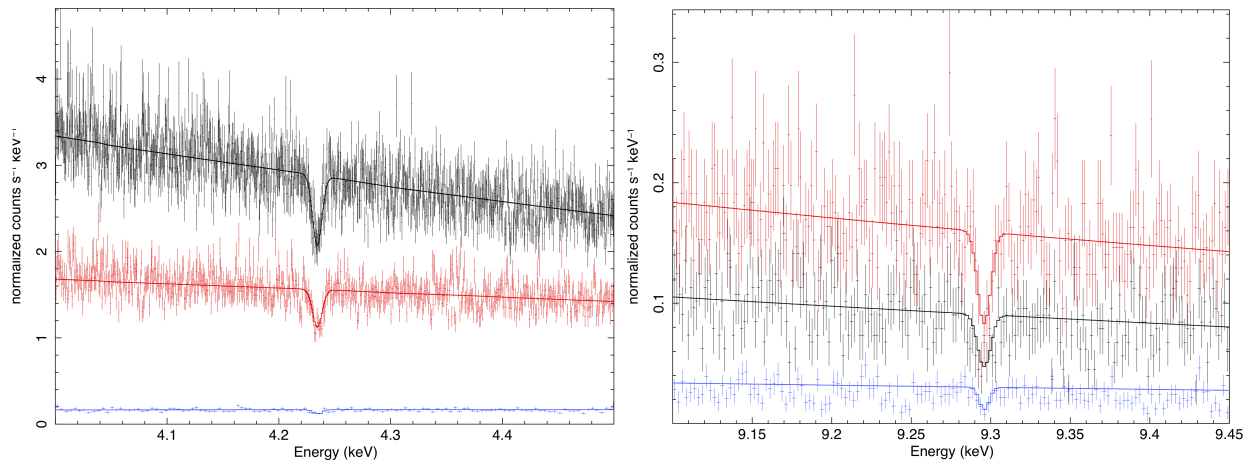

  \centering
  \vspace{-0.5cm}
  \includegraphics[width=0.49\textwidth]{hitomi_upper.png}
  \includegraphics[width=0.5\textwidth]{hitomi_lower.png}
 %
    \caption{Simulations of the lines detected with Hitomi (blue) from PSR J1833-1034 for TES-based telescope for the same observing time with two mirror configurations: single-bounce collector (black) with three times the geometric area of NICER and double-bounce collector (red) with the same geometric area as NICER.}
    \label{fig:rev}
\end{figure}
Many neutron stars are also rapidly rotating. Observing phase-resolved spectral features adds the requirement of high-time-resolution to high-throughput. 
Rotation imparts a particular pattern in the observed X-rays as a function of energy and phase. In particular, if only a portion of the surface is emitting, hard X-rays will lead softer ones \cite{1999ApJ...519L..73F}, and if the emission pattern is known or can be constrained from observations as it could be in Type-I X-ray burst oscillations \cite{2005MNRAS.361..504H}, one can constrain the mass and radius of the neutron star \cite{0004-637X-787-2-136,0004-637X-832-1-92}.  The boost in effective area of large collector based telescope with TES arrays relative to \textit{NICER} will allow us to study fainter objects in shorter times and to derive constraints from ensembles of Type-I X-ray bursts with oscillations. The dramatic increase in energy resolution will probe and constrain the underlying emission models, reducing potential systematic errors in the determinations of mass and radius.

Recently, the Hitomi satellite found strong evidence for weak and narrow absorption lines from the rotational-powered pulsar PSR~J1833-1034 in the supernova remnant G21.5-0.9 \cite{2018PASJ..tmp...37A} at 4.2345~keV and 9.296~keV.  The observation is presented in Fig.~\ref{fig:rev} in blue. The red and black points show the simulated results with TES arrays coupled to collector optics.  Depending on which line and on the configuration of the telescope, such a configuration would find two to ten times more photons within the line than Hitomi, with similar exposure time.  Such an instrument would also open the possibility of phase-resolved spectroscopy to verify that the feature indeed originiates from the pulsar rather than the supernova remnant.  PSR~J1833-1034 rotates with a period of 66~ms, so the rotational velocity at the equator of the neutron star is about 1000~km/s, yielding a width of about 15~eV, just a factor of two larger than Hitomi's energy resolution.  If the line indeed originates at the surface, it is somewhat surprising that Hitomi discovered such a narrow feature.

The rapidly rotating (3.15~ms) and massive ($1.97\pm0.04 \mathrm{M}_\odot$) pulsar PSR~J1614-2230 is an excellent target \cite{2016ApJ...822...27M} for high-time-resolution spectroscopy. Pulsations have been detected by NICER \cite{2019AAS...23315313W}, however it is somewhat too faint for a long targeted NICER observation that would constrain the neutron star's radius from a detailed analysis of its pulse profile. Observing PSR~J1614-2230 with greater effective area
could yield powerful constraints on the radius of this neutron star as well as the emission mechanism to control systematics. Given that this is a faint source, pile-up is not an issue.  Because the field of this star is weak and it has not recently accreted, we do not expect to see spectral features, so the high-energy resolution is not crucial. 

Strong evidence for spectral lines has been found for several slowly rotating neutron stars with stronger magnetic fields. The XDIN RX~J1308.6+2127 exhibits a spectral feature at about 740~eV with an equivalent width of about 15~eV \cite{2017MNRAS.468.2975B} over only a portion of its rotation.  Unfortunately, the energy resolution of the EPIC-pn instrument on XMM-Newton is insufficient to resolve the line.  A comprehensive analysis of the available data for the XDINS with XMM-Newton yields upper limits on the equivalent width of narrow (unresolved lines) of 10-50~eV, depending on the source and the duration of the available observations.  A TES spectrometer (with energy resolution of about 1~eV) with a similar effective area to XMM-Newton would yield constraints ten times stronger for similar observing times.   
  \begin{figure}[tb]
      \centering
      \includegraphics[width=0.9\textwidth]{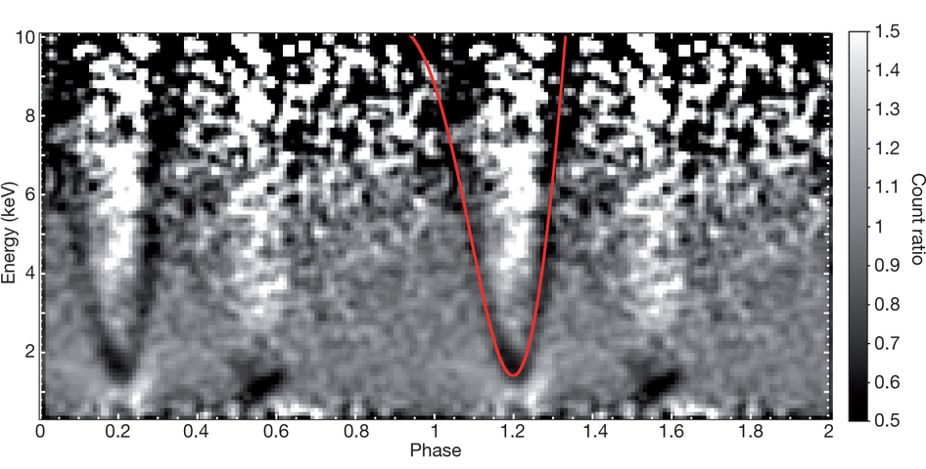}
      \caption{Phase resolved spectra from SGR~0418+5729 with the EPIC instrument on XMM-Newton.  The red curve depicts the line centroid from a model in which radiation from the surface is absorbed by protons through the cyclotron resonance in a baryon-loaded current loop above the surface of the neutron star. Figure from \cite{2013Natur.500..312T}.}
      \label{fig:sgr}
  \end{figure}  
  
  Magnetars are among the most magnetic compact objects in the Universe. Their high energy properties and spin parameters point to a super-strong magnetic field of the order of $10^{14}-10^{15}$~G. Spectroscopy provides a direct diagnostic of their total surface magnetic field strength. While their electron cyclotron features would fall in the MeV band, their proton cyclotron lines fall in the X-ray band. To date, we have evidence, initially mostly from the RXTE satellite and more recently from just a few observations with operating X-ray missions, of sporadic detections of spectral features in magnetars X-ray spectra. While the interpretation of these lines is still being debated, and their detection occurred either during an outburst or in quiescence, they have been mostly interpreted as proton cyclotron features from a magnetar-strength magnetic field, confirming in many cases the high magnetic field value inferred from spin-down measurements (e.g., 5 keV absorption line from SGR1806-20~\cite{2002ApJ...574L..51I}; 8.1 keV absorption line from 1RXS J170849-4009104~\cite{2004NuPhS.132..554R};  4 keV and 8 keV emission lines from 4U 0142+62~\cite{2008AIPC..983..234G}). More recently a variable absorption feature near 2 keV was discovered in a phase-resolved spectroscopy (Fig.~\ref{fig:sgr}) of the magnetar SGR~0418+5729 whose spin properties point to a much lower, below the QED value, magnetic field ($6\times 10^{12}$G),  supporting high-order multipolar field components \cite{2013Natur.500..312T}.   The line, when interpreted as a proton cyclotron feature, yields  a magnetic field ranging from $2 \times 10^{14}-10^{15}$~G. This suggests that spectroscopy can directly probe the topology of the magnetic field, and in ways that can not be done with timing which infers the dipole field strength.   Unfortunately, the XMM-Newton EPIC in-struments have insufficient energy resolution to resolve the feature. Furthermore, the data depicted in Fig.~\ref{fig:sgr} hint of a second feature on the opposite hemisphere of the star. Observing similar lines in more neutron stars and with higher sensitivity could reveal the structure of the magnetic field and how magnetars work.

While early theoretical predictions suggested relatively wide absorption lines \cite[e.g.][]{2001ApJ...560..384Z,2001MNRAS.327.1081H} as observed in some of the magnetar bursts’ spectra, vacuum polarization has been subsequently suggested to suppress the strength of the proton cyclotron resonances in strongly magnetized plasma  \cite{2002ApJ...566..373L,2003ApJ...583..402O}. This could reduce the line equivalent width by nearly an order of magnitude. 
TES arrays will open a new window for a higher sensitivity search for the proton or ion cyclotron features (or atomic lines from high Z elements) with a weak (shallow or narrow) line, and will be especially suited to studying bright burst spectra as well as monitor the evolution of magnetars’ spectra. 



If we turn our focus to accreting neutron stars, the highest-frequency quasi-periodic oscillations (QPOs) observed in accreting neutron star systems can provide unique constraints on the neutron stars themselves \cite{1998ApJ...508..791M} if the oscillation can be associated with motion near the inner edge of the accretion disk.  
The current record is 4U~0614+09, which has a QPO with three-sigma lower limit on its frequency of 1267~Hz, yielding a constraint on the mass of this object of less than 2.1~M$_\odot$ \cite{2018MNRAS.479..426V}.  High-resolution spectroscopy of the QPO itself can bring the power of these constraints forward to obtain stellar mass measurements and probe the spacetime around the neutron star as well (perhaps measuring the moment of inertia).  The objects that have been found to exhibit these high frequency QPOs are typically brighter than 0.1~Crab (this is in part a selection effect).  They are sufficiently bright that a focusing instrument like the \textit{X-IFU} on \textit{ATHENA} \cite{2016SPIE.9905E..2FB} cannot perform spectroscopy due to photon pile-up, but a TES experiment with collecting optics could provide exciting measurements of neutron star masses and moments of inertia, as well as a basic test of the models for the QPOs.

%
%

\clearpage
\bibliographystyle{jer}
\bibliography{main}

\begin{thebibliography}{36}
\providecommand{\natexlab}[1]{#1}
\providecommand{\url}[1]{\texttt{#1}}
\expandafter\ifx\csname urlstyle\endcsname\relax
  \providecommand{\doi}[1]{doi: #1}\else
  \providecommand{\doi}{doi: \begingroup \urlstyle{rm}\Url}\fi

\bibitem[{Steiner} et~al.(2013){Steiner}, {Lattimer}, and
  {Brown}]{2013ApJ...765L...5S}
A.~W. {Steiner}, J.~M. {Lattimer}, and E.~F. {Brown}.
\newblock \emph{\apjl}, 765:\penalty0 L5, 2013.

\bibitem[Morgan et~al.(2017)Morgan, Pappas, Bennett, Gard, Hays-Wehle, Hilton,
  Reintsema, Schmidt, Ullom, and Swetz]{doi:10.1063/1.4984065}
K.~M. Morgan, C.~G. Pappas, D.~A. Bennett, J.~D. Gard, J.~P. Hays-Wehle, G.~C.
  Hilton, C.~D. Reintsema, D.~R. Schmidt, J.~N. Ullom, and D.~S. Swetz.
\newblock \emph{Applied Physics Letters}, 110\penalty0 (21):\penalty0 212602,
  2017.

\bibitem[Morgan et~al.(2016)Morgan, Alpert, Bennett, Denison, Doriese, Fowler,
  Gard, Hilton, Irwin, Joe, O'Neil, Reintsema, Schmidt, Ullom, and
  Swetz]{doi:10.1063/1.4962636}
K.~M. Morgan, B.~K. Alpert, D.~A. Bennett, E.~V. Denison, W.~B. Doriese, J.~W.
  Fowler, J.~D. Gard, G.~C. Hilton, K.~D. Irwin, Y.~I. Joe, G.~C. O'Neil, C.~D.
  Reintsema, D.~R. Schmidt, J.~N. Ullom, and D.~S. Swetz.
\newblock \emph{Applied Physics Letters}, 109\penalty0 (11):\penalty0 112604,
  2016.

\bibitem[{Ho} et~al.(2007){Ho}, {Kaplan}, {Chang}, {van Adelsberg}, and
  {Potekhin}]{2007MNRAS.375..821H}
W.~C.~G. {Ho}, D.~L. {Kaplan}, P.~{Chang}, M.~{van Adelsberg}, and A.~Y.
  {Potekhin}.
\newblock \emph{\mnras}, 375:\penalty0 821--830, 2007.

\bibitem[{{\"O}zel} and {Psaltis}(2003)]{2003ApJ...582L..31O}
F.~{{\"O}zel} and D.~{Psaltis}.
\newblock \emph{\apj}, 582:\penalty0 L31--L34, 2003.

\bibitem[{Baub{\"o}ck} et~al.(2013){Baub{\"o}ck}, {Psaltis}, and
  {{\"O}zel}]{2013ApJ...766...87B}
M.~{Baub{\"o}ck}, D.~{Psaltis}, and F.~{{\"O}zel}.
\newblock \emph{\apj}, 766:\penalty0 87, 2013.

\bibitem[{Strohmayer} and {Markwardt}(2010)]{2010ATel.2929....1S}
T.~E. {Strohmayer} and C.~B. {Markwardt}.
\newblock \emph{The Astronomer's Telegram}, 2929, 2010.

\bibitem[{Yoneda} et~al.(2018){Yoneda}, {Done}, {Paerels}, {Takahashi}, and
  {Watanabe}]{2018MNRAS.475.2194Y}
H.~{Yoneda}, C.~{Done}, F.~{Paerels}, T.~{Takahashi}, and S.~{Watanabe}.
\newblock \emph{\mnras}, 475:\penalty0 2194--2203, 2018.

\bibitem[{Cottam} et~al.(2002){Cottam}, {Paerels}, and
  {Mendez}]{2002Natur.420...51C}
J.~{Cottam}, F.~{Paerels}, and M.~{Mendez}.
\newblock \emph{\nat}, 420:\penalty0 51--54, 2002.

\bibitem[{Galloway} et~al.(2010){Galloway}, {Lin}, {Chakrabarty}, and
  {Hartman}]{2010ApJ...711L.148G}
D.~K. {Galloway}, J.~{Lin}, D.~{Chakrabarty}, and J.~M. {Hartman}.
\newblock \emph{\apjl}, 711:\penalty0 L148--L151, 2010.

\bibitem[{Lin} et~al.(2010){Lin}, {{\"O}zel}, {Chakrabarty}, and
  {Psaltis}]{2010ApJ...723.1053L}
J.~{Lin}, F.~{{\"O}zel}, D.~{Chakrabarty}, and D.~{Psaltis}.
\newblock \emph{\apj}, 723:\penalty0 1053--1056, 2010.

\bibitem[{Weinberg} et~al.(2006){Weinberg}, {Bildsten}, and
  {Schatz}]{2006ApJ...639.1018W}
N.~N. {Weinberg}, L.~{Bildsten}, and H.~{Schatz}.
\newblock \emph{\apj}, 639:\penalty0 1018--1032, 2006.

\bibitem[{Kajava} et~al.(2017){Kajava}, {N{\"a}ttil{\"a}}, {Poutanen},
  {Cumming}, {Suleimanov}, and {Kuulkers}]{2017MNRAS.464L...6K}
J.~J.~E. {Kajava}, J.~{N{\"a}ttil{\"a}}, J.~{Poutanen}, A.~{Cumming},
  V.~{Suleimanov}, and E.~{Kuulkers}.
\newblock \emph{\mnras}, 464:\penalty0 L6--L10, 2017.

\bibitem[{Li} et~al.(2018){Li}, {Suleimanov}, {Poutanen}, {Salmi}, {Falanga},
  {N{\"a}ttil{\"a}}, and {Xu}]{2018ApJ...866...53L}
Z.~{Li}, V.~F. {Suleimanov}, J.~{Poutanen}, T.~{Salmi}, M.~{Falanga},
  J.~{N{\"a}ttil{\"a}}, and R.~{Xu}.
\newblock \emph{\apj}, 866:\penalty0 53, 2018.

\bibitem[{Barret} et~al.(2016)]{2016SPIE.9905E..2FB}
D.~{Barret} et~al.
\newblock In \emph{Space Telescopes and Instrumentation 2016: Ultraviolet to
  Gamma Ray}, volume 9905 of \emph{\procspie}, page 99052F, 2016.

\bibitem[{Galloway} et~al.(2008){Galloway}, {Muno}, {Hartman}, {Psaltis}, and
  {Chakrabarty}]{2008ApJS..179..360G}
D.~K. {Galloway}, M.~P. {Muno}, J.~M. {Hartman}, D.~{Psaltis}, and
  D.~{Chakrabarty}.
\newblock \emph{\apjs}, 179:\penalty0 360-422, 2008.

\bibitem[{Rutledge} et~al.(2002){Rutledge}, {Bildsten}, {Brown}, {Pavlov}, and
  {Zavlin}]{2002ApJ...577..346R}
R.~E. {Rutledge}, L.~{Bildsten}, E.~F. {Brown}, G.~G. {Pavlov}, and V.~E.
  {Zavlin}.
\newblock \emph{\apj}, 577:\penalty0 346--358, 2002.

\bibitem[{Wijnands} et~al.(2017){Wijnands}, {Degenaar}, and
  {Page}]{2017JApA...38...49W}
R.~{Wijnands}, N.~{Degenaar}, and D.~{Page}.
\newblock \emph{Journal of Astrophysics and Astronomy}, 38:\penalty0 49, 2017.

\bibitem[{Ford}(1999)]{1999ApJ...519L..73F}
E.~C. {Ford}.
\newblock \emph{\apjl}, 519:\penalty0 L73--L75, 1999.

\bibitem[{Heyl}(2005)]{2005MNRAS.361..504H}
J.~S. {Heyl}.
\newblock \emph{\mnras}, 361:\penalty0 504--510, 2005.

\bibitem[Psaltis et~al.(2014)Psaltis, Ozel, and
  Chakrabarty]{0004-637X-787-2-136}
D.~Psaltis, F.~Ozel, and D.~Chakrabarty.
\newblock \emph{The Astrophysical Journal}, 787\penalty0 (2):\penalty0 136,
  2014.

\bibitem[Ozel et~al.(2016)Ozel, Psaltis, Arzoumanian, Morsink, and
  Baubock]{0004-637X-832-1-92}
F.~Ozel, D.~Psaltis, Z.~Arzoumanian, S.~Morsink, and M.~Baubock.
\newblock \emph{The Astrophysical Journal}, 832\penalty0 (1):\penalty0 92,
  2016.

\bibitem[{Hitomi Collaboration} et~al.(2018)]{2018PASJ..tmp...37A}
{Hitomi Collaboration} et~al.
\newblock \emph{\pasj}, 70:\penalty0 38, 2018.

\bibitem[{Miller}(2016)]{2016ApJ...822...27M}
M.~C. {Miller}.
\newblock \emph{\apj}, 822:\penalty0 27, 2016.

\bibitem[{Wolff} et~al.(2019){Wolff}, {Guillot}, {Ho}, and
  {Ray}]{2019AAS...23315313W}
M.~{Wolff}, S.~{Guillot}, W.~C. {Ho}, and P.~S. {Ray}.
\newblock In \emph{American Astronomical Society Meeting Abstracts \#233},
  volume 233 of \emph{American Astronomical Society Meeting Abstracts}, page
  153.13, 2019.

\bibitem[{Borghese} et~al.(2017){Borghese}, {Rea}, {Coti Zelati}, {Tiengo},
  {Turolla}, and {Zane}]{2017MNRAS.468.2975B}
A.~{Borghese}, N.~{Rea}, F.~{Coti Zelati}, A.~{Tiengo}, R.~{Turolla}, and
  S.~{Zane}.
\newblock \emph{\mnras}, 468:\penalty0 2975--2983, 2017.

\bibitem[{Tiengo} et~al.(2013){Tiengo}, {Esposito}, {Mereghetti}, {Turolla},
  {Nobili}, {Gastaldello}, {G{\"o}tz}, {Israel}, {Rea}, {Stella}, {Zane}, and
  {Bignami}]{2013Natur.500..312T}
A.~{Tiengo}, P.~{Esposito}, S.~{Mereghetti}, R.~{Turolla}, L.~{Nobili},
  F.~{Gastaldello}, D.~{G{\"o}tz}, G.~L. {Israel}, N.~{Rea}, L.~{Stella},
  S.~{Zane}, and G.~F. {Bignami}.
\newblock \emph{\nat}, 500:\penalty0 312--314, 2013.

\bibitem[{Ibrahim} et~al.(2002){Ibrahim}, {Safi-Harb}, {Swank}, {Parke},
  {Zane}, and {Turolla}]{2002ApJ...574L..51I}
A.~I. {Ibrahim}, S.~{Safi-Harb}, J.~H. {Swank}, W.~{Parke}, S.~{Zane}, and
  R.~{Turolla}.
\newblock \emph{\apj}, 574:\penalty0 L51--L55, 2002.

\bibitem[{Rea} et~al.(2004){Rea}, {Israel}, and {Stella}]{2004NuPhS.132..554R}
N.~{Rea}, G.~L. {Israel}, and L.~{Stella}.
\newblock \emph{Nuclear Physics B Proceedings Supplements}, 132:\penalty0
  554--559, 2004.

\bibitem[{Gavriil} et~al.(2008){Gavriil}, {Dib}, and
  {Kaspi}]{2008AIPC..983..234G}
F.~P. {Gavriil}, R.~{Dib}, and V.~M. {Kaspi}.
\newblock In C.~{Bassa}, Z.~{Wang}, A.~{Cumming}, and V.~M. {Kaspi}, editors,
  \emph{40 Years of Pulsars: Millisecond Pulsars, Magnetars and More}, volume
  983 of \emph{American Institute of Physics Conference Series}, pages
  234--238, 2008.

\bibitem[{Zane} et~al.(2001){Zane}, {Turolla}, {Stella}, and
  {Treves}]{2001ApJ...560..384Z}
S.~{Zane}, R.~{Turolla}, L.~{Stella}, and A.~{Treves}.
\newblock \emph{\apj}, 560:\penalty0 384--389, 2001.

\bibitem[{Ho} and {Lai}(2001)]{2001MNRAS.327.1081H}
W.~C.~G. {Ho} and D.~{Lai}.
\newblock \emph{\mnras}, 327:\penalty0 1081--1096, 2001.

\bibitem[{Lai} and {Ho}(2002)]{2002ApJ...566..373L}
D.~{Lai} and W.~C.~G. {Ho}.
\newblock \emph{\apj}, 566:\penalty0 373--377, 2002.

\bibitem[{{\"O}zel}(2003)]{2003ApJ...583..402O}
F.~{{\"O}zel}.
\newblock \emph{\apj}, 583:\penalty0 402--409, 2003.

\bibitem[{Miller} et~al.(1998){Miller}, {Lamb}, and
  {Psaltis}]{1998ApJ...508..791M}
M.~C. {Miller}, F.~K. {Lamb}, and D.~{Psaltis}.
\newblock \emph{\apj}, 508:\penalty0 791--830, 1998.

\bibitem[{van Doesburgh} et~al.(2018){van Doesburgh}, {van der Klis}, and
  {Morsink}]{2018MNRAS.479..426V}
M.~{van Doesburgh}, M.~{van der Klis}, and S.~M. {Morsink}.
\newblock \emph{\mnras}, 479:\penalty0 426--434, 2018.

\end{thebibliography}

\end{document}